\documentstyle[prd,aps,epsfig]{revtex}
\tighten 
\begin{document} 
\def\teuton{\cal} 
\draft \twocolumn[\hsize\textwidth\columnwidth\hsize\csname 
@twocolumnfalse\endcsname 
\title{Twin Paradox and Space Topology} 
\author{Jean--Philippe Uzan$^{1,2}$, Jean--Pierre Luminet$^3$, Roland 
Lehoucq$^4$ 
        and Patrick Peter$^{3,5}$} 
\address{(1) Laboratoire de Physique Th\'eorique, CNRS--UMR~8627, 
B\^at. 210, Universit\'e Paris XI, F-91405 Orsay cedex (France)\\ 
         (2) D\'epartement de Physique Th\'eorique, Universit\'e de 
Gen\`eve, 24 quai E. Ansermet, CH-1211 Gen\`eve 4 (Switzerland)\\ 
         (3) D\'epartement d'Astrophysique Relativiste et de 
Cosmologie, Observatoire de Paris, CNRS--UMR~8629,\\ F--92195 
Meudon cedex (France)\\ 
         (4) CE--Saclay, DSM/DAPNIA/Service d'Astrophysique, F--91191 
Gif sur Yvette cedex (France).\\
         (4) Institut d'Astrophysique de Paris, 98 bis, boulevard 
Arago, 75005 Paris (France)} 
\date{\today} 
\maketitle 
\begin{abstract} 
If space is compact, then a traveller twin can leave Earth, travel  
back home without changing direction and find her sedentary twin older  
than herself.  We show that the asymmetry between their spacetime  
trajectories lies in a topological invariant of their spatial  
geodesics, namely the homotopy class.  This illustrates how the  
spacetime symmetry invariance group, although valid {\it locally}, is  
broken down {\it globally} as soon as some points of space are  
identified.  As a consequence, any non--trivial space topology defines  
preferred inertial frames along which the proper time is longer than  
along any other one. 
 
\end{abstract} 
 
\pacs{{\bf Preprint numbers:} LPT--ORSAY 00/45, 
UGVA--DPT 00/04--1080} \vskip0.5cm ] 
%%%%%%%%%%%%%%%%%%%%%%%%%%%%%%%%%%%%%%%%%%%%%%%%%%%%%%%%%%%%%%%%%%%%%%% 
 
\section{Introduction} 
 
The twin paradox is the best known thought experiment of special  
relativity, whose resolution provides interesting insights on the  
structure of spacetime and on the applicability of the Lorentz  
transformations.  In its seminal paper on special relativity, Einstein  
\cite{einstein05} pointed out the problem of clocks synchronization  
between two inertial frames with relative velocity $v$.  Later on,  
Langevin \cite{langevin11} picturesquely formulated the problem by  
taking the example of twins aging differently according to their  
respective worldlines.  The keypoint for understanding the paradox is  
the asymmetry between the spacetime trajectories of the ``sedentary  
twin'' and of the ``traveller twin''.  The subject has been widely  
studied for pedagogical purpose \cite{romer59,taywheel92}, the role of  
acceleration was examined in details  
\cite{unruh81,nikolic99,good82,boughn89} and a full general  
relativistic treatment was given \cite{perrin70}. 
 
Although counter-intuitive, the twin paradox is clearly not a logical  
contradiction, it merely illustrates the elasticity of time in  
relativistic mechanics.  The experiment was actually performed in 1971  
with twin atomic clocks initially synchronized, one of them being kept  at rest on Earth and the other one being embarked on a commercial  
flight: the time shifts perfectly agreed with the fully relativistic  
calculations \cite{hafele} 
 
An interesting ``revisited" paradox was formulated  
\cite{brans73,low90} in the framework of a closed space (due to  
curvature or to topology). In such a case, the twins can meet again  
without none of them being accelerated, yet they aged differently.   
Both an algebraic and a geometric solution were given \cite{dray90}. 
  
Our present goal is to extend such explanations by adding a topological  
characterization of reference frames, which allows us to solve the  
twin paradox {\it whatever} the global shape of space may be.   
We first briefly recall, in \S~\ref{II}, the classical twin paradox  
and its standard resolution, in \S~\ref{III} we investigate the case  
of a spacetime with compact spatial sections and in \S~\ref{IV}, we  
show that the root explanation of the twin paradox lies in the global  
breakdown of the spacetime symmetry group by a non--trivial topology. 
 
\section{The standard twin paradox}\label{II} 
 
Let observers $1$ and $2$ be attached to inertial frames with relative 
velocity $v$. $1$ is supposed to be at rest and to experience no acceleration. At time 
$t=0$, the observers synchronise their clock (thus they can be  
called ``twins"). Then twin $2$ travels away at 
velocity $+v$ with respect to $1$ and comes back with velocity 
$-v$.  According to special relativity, the travel time, $\Delta\tau_2$, 
measured by $2$ (its proper time) is related to the proper time measured by 
$1$, $\Delta\tau_1$, by 
\begin{equation}\label{1} 
\Delta\tau_2=\sqrt{1-v^2}\Delta\tau_1. 
\end{equation} 
A paradox arises if one considers that the situation is perfectly  
symmetrical about $1$ and $2$, since $2$ sees $1$ travelling away with velocity  
$-v$ and coming back with velocity $+v$. If this was correct, one could  
reverse the reasoning to deduce that $1$ should be younger than $2$,  
with 
\begin{equation}\label{2} 
\Delta\tau_1=\sqrt{1-v^2}\Delta\tau_2, 
\end{equation} 
and an obvious contradiction would arise. 
 
Indeed, the  
previous argument holds whenever $2$ is not accelerating. As first  
explained by Paul Langevin~\cite{langevin11}, among all the worldlines that connect two  
spacetime events (such as the departure and return of $2$), the one  
which has the longest proper time is the unaccelerated one, i.e.  the  
reference frame ${\bf K}$ of $1$.  The traveller  
twin $2$ cannot avoid accelerating and decelerating to make its return  
journey; then he had to jump from an inertial frame ${\bf K}'$ moving  
relatively to ${\bf K}$ with velocity $v$ to {\it another} inertial  
frame ${\bf K}''$ moving with velocity $-v$ with respect to ${\bf K}$.   
Hence the situation is not symmetrical about the twins: a kink  
(infinite acceleration) in  
the middle of the path of twin $2$ explains the difference, and there is  
no contradiction in the fact that the sedentary twin $1$ will definitely  
be older than the traveller twin $2$.  
 
The same result holds in the framework of general relativity, dealing with  
a more realistic situation including accelerations, gravitational  
fields and curved spacetime (so that the kink is smoothed out): in  
order to achieve its journey, the traveller $2$ necessarily  
experiences a finite and variable acceleration; thus her reference  
frame is not equivalent to that of $1$. 
 
Such explanations, as rephrased by Bondi~\cite{bondi64}, are  
equivalent to say that {\it there is only one way of getting from the  
first meeting point to the second without acceleration}.  
 
However, acceleration is not the only and essential point of the twin  
paradox, as shown by the example of non--accelerated twins in a closed  
space, in which there are several ways to go from the first meeting  
point to the second one {\it without accelerating}  
\cite{brans73,low90,dray90}.  The key explanation of the twin paradox  
is now ``some kind" of asymmetry between the spacetime paths joining  
two events.  We investigate below the nature of such an asymmetry when  
space topology is {\it not} trivial (i.e.  simply--connected)

\section{Twins in a compact space}\label{III}

In a spacetime which has at least one compact space dimension, one can  
actually start from one point, travel along several straight geodesics  
and come back to the same spatial position without accelerating nor  
decelerating.  Einstein's relativity theory determines the local  
properties of the spacetime ${\cal M}$ (its metrics), but gives little  
information about its global properties (its topology)  
\cite{geroch71,luminet95}, so that special relativity (in the absence  
of gravitational fields) or general relativity (involving  
gravitational fields in curved spacetimes) well describe the local  
physics.  For instance, the Minkowski spacetime (${\cal M}, \eta$)  
used in special relativity is a manifold ${\bf R}^4$ with a flat  
Lorentzian metric $\eta$ = diag(-1,1,1,1) and Euclidean space sections  
$\Sigma$.  One can obain spacetimes locally identical to (${\cal M},  
\eta$) but with different large scale properties by identifying points  
in $\cal M$ under a group of transformations, called {\it holonomies},  
which preserve the metric (thus holonomies are isometries).  For instance, identifying ($x_0, x_1, x_2,  
x_4$) with ($x_0+L, x_1, x_2, x_3$) changes the topology from ${\bf  
R}^4$ to a cylinder ${\bf S}^1\times{\bf R}^3$ and introduces closed  
timelike lines.  However, if causality is believed to hold in the  
sense that no effect can preceed its cause, such an identification is  
prohibited, and the study of spacetime topology is restricted so as to  
exclude closed timelike curves \cite{geroch71}.  This is achieved if  
spacetime can be decomposed as the direct product 
\begin{equation} 
{\cal M}={\bf R}\times\Sigma 
\end{equation} 
where the real axis ${\bf R}$ refers to the time direction and  
$\Sigma$ to the three dimensional spatial sections.  Now the topology  
of spacetime amounts to the study of the various shapes of the  
spatial sections $\Sigma$.   
 
The topology of a three--dimensional Riemannian space $\Sigma$ can be described  
in full generality by a fundamental polyhedron ${\cal P}$ and a  
holonomy group $\Gamma$ whose elements $g$ identify the faces of the  
polyhedron by pairs (see e.g.  \cite{luminet95} for a general  
discussion of the topological properties of spacetimes).  It follows  
that $\Sigma$ can be written as 
\begin{equation} 
\Sigma=X/\Gamma 
\end{equation} 
where $X$ is the universal covering space \cite{luminet95} 
(simply--connected and three dimensional) 
and the quotient refers to the equivalence relation `$\equiv$' defined as 
\begin{equation}\label{id} 
\forall\,{\bf x},{\bf y}\in X\quad 
{\bf x}\equiv{\bf y}\Longleftrightarrow \left(\exists g\in\Gamma\quad|\quad  
{\bf x}=g{\bf y}\right). 
\end{equation} 
 
If $\Gamma$ reduces to the identity, space is simply--connected, in  
the sense that two points of space are connected by only one geodesic.   
As soon as there are non--trivial holonomies which identify points,  
space is multi--connected and {\it several} geodesics connect two any  
distinct points.  In a cosmological context, multi--connected universe  
models lead to the existence of ghost images in the observable  
universe when one topological length is shorter than the horizon size.   
Many methods aimed to detect the cosmic topology have been proposed  
\cite{luminet95,topobib}, so far with no definite answer coming from  
observational data. 
 
\begin{figure} 
\centering{ 
\epsfig{figure=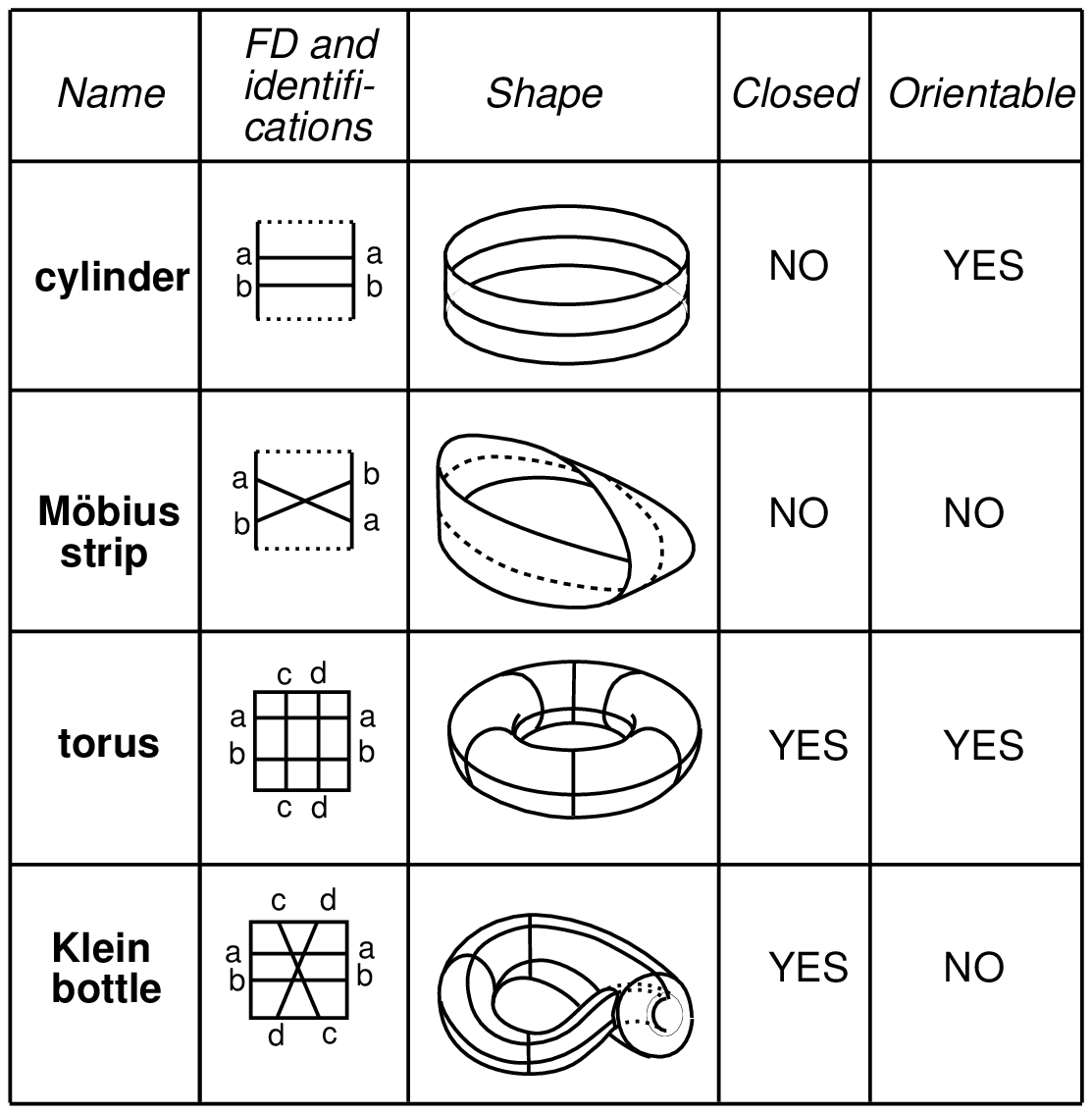, width=6cm} 
\vskip0.25cm 
\caption{The four multi--connected topologies of the two dimensional 
Euclidean plane. They are constructed from a 
parallelogram or an infinite band (fundamental domain FD), by  
identification of edges according to the allowable 
holonomies. We indicate as well their compactness and orientability  
properties (from [14]).} 
\label{fig1}} 
\end{figure}

Returning to Minkowski space for the sake of clarity, the holonomies  
of $\Gamma$ that preserve the flatness of the space sections $\Sigma$  
are the identity, the translations, the reflexions and the helicoidal  
displacements.  The group classification leads to 18 Euclidean  
spaceforms with different topologies, all of them having as universal  
covering space $X$ the simply--connected, infinite Euclidean space  
${\bf R}^3$.  Six of them are compact (i.e.  of finite volume) and  
orientable.  Their fundamental polyhedron can be a parallelepiped or  
a hexagonal prism.  For the sake of visualization, in the following  
we shall develop our reasoning in flat spacetimes with $1+2$  dimensions only, i.e.  whose spatial sections $\Sigma$ are just  
surfaces.  In such a case there are 5 Euclidean topologies: the  
cylinder, the M\"obius strip, the flat torus, the Klein bottle and the  
Euclidean plane \cite{luminet95}.  For a pedagogical purpose, we select  
the case where space has a torus--like topology (see  
figure~\ref{fig1}).  However we emphasize that our conclusions will  
remain valid in (1+3) dimensions, whatever the topology and the  
(constant) space curvature may be. 
 
\begin{figure} 
\centering{ 
\epsfig{figure=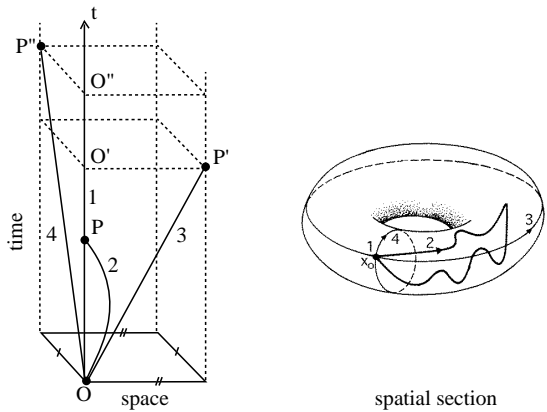, width=8cm} 
\vskip0.25cm 
\caption{Different twins in a (1+2) spacetime with toroidal spatial 
sections. All of them leave $0$ at the same time. While $1$ remains at 
home, $2$ [dash-dotted line] goes away and then comes back to meet $1$ in 
$P$ (it corresponds to the standard case), $3$ [solid line] goes around 
the universe in a given direction from $O$ to $P' \equiv O'$ and $4$ 
[dashed line] goes around 
the universe  along another direction from $O$ to 
$P'' \equiv O''$.  On the left plot, we used the fundamental domain 
representation (in which the 2--torus corresponds to a rectangle 
whose edges $/$ and $//$ are respectively identified; see 
figure~\ref{fig1}). On the right plot, we depict the projection of their 
trajectory on a constant time slice. The spacetime points $O$, $O'$,  
$O''$, $P$, 
$P'$ and $P''$ of ${\cal M}$ are projected onto the same base point ``1'' in 
$\Sigma$.} 
\label{fig2}} 
\end{figure}

Since a characteristics of multi--connected spaces is to provide  
several geodesics between two any points, we shall extend the concept  
of ``twins" to more than two and consider an ensemble of ``twins"  
(strictly speaking, initially synchronized clocks)  
labeled as 1,2,3 and 4 (see figure~\ref{fig2} [left]). The twin $1$ stays at home at point $O$ and  
her worldline 
can be identified with the time axis.  The twin 2 leaves $O$ at $t=0$, 
travels away and then turns back to meet twin 1 in $P$. The twins 
$3$ and $4$ leave $O$ in two different directions along  
non--accelerated worldlines and travel away from 
$1$; they respectively reach $P'$ and $P''$, where they 
meet twin $1$\footnote{Under very special circumstances, twins $3$  
and $4$ could also meet altogether with twins $1$ and $2$ at the same  
point of spacetime!}. 
 Due to warped space, 
twins $3$ and $4$ managed to come back without changing their direction  
and inertial frame. Now, one wants to compare the ages within the various  
pairs  
of twins when they meet again. Whereas twin $2$ 
undergone the standard paradox, there seems to be a real paradox with  
twins $1$, $3$ and $4$, who all followed strictly inertial frames.   
 
In \cite{dray90} it was shown that, in the case of a cylinder, the  
sedentary twin $1$ is always older than any traveller because their  
states of motions, although non--accelerated, are not symmetrical.   
Which kind of asymmetry is to be considered?  As we emphasize below,  
acceleration is a local concept, and the only explanation lies in a  
global breakdown of symmetry due to a non--trivial topology. 
 
Let us consider the projection of the 
worldlines onto a constant time hypersurface $\Sigma$,  assumed 
to be a 2D flat torus (see figure~\ref{fig2} [right]). Each projection is a 
loop $\gamma(u,{\bf x}_0)$ at ${\bf x}_0$ which can be parametrised by 
$u\in[0,1[$ if $\gamma(0)=\gamma(1)={\bf x}_0$, where ${\bf x}_0$ is a 
point of $\Sigma$.  Two loops at ${\bf x}_0$, $\gamma$ and $\delta$, are 
said to be homotopic ($\gamma\sim\delta$) if they can be {\it 
continuously} deformed into one another, i.e. if there exists a  
continuous map (called homotopy) $F:[0,1[\times[0,1[\rightarrow X$ such that 
$$ 
\forall u\in[0,1[,\quad F(s,0)=\gamma(u),\,F(u,1)=\delta(u) 
$$ 
$$ 
\forall v\in[0,1[,\quad F(0,v)=F(1,v)={\bf x}_0. 
$$ 
The equivalence class of homotopic loops is denoted by $[\gamma]$.  We  
denote $\gamma_i$ the loop corresponding to the projection of the  
trajectory of twin $i$ in $\Sigma$. 
 
In our example (see figure~\ref{fig2}), the twins $1$ and $2$ have  
homotopic trajectories: since $2$ does not ``go around'' the universe,  
the loop $\gamma_2$ can be continuously contracted into the null loop  
$\gamma_1=\lbrace0\rbrace$, so 
$$\gamma_1\sim\gamma_2\sim\lbrace0\rbrace.$$ 
  However, among these two  
homotopic loops, only one corresponds to an inertial  
observer going from $O$ to $P$: that of twin $1$, which is thus older  
than twin $2$, as  
expected in the standard paradox.  
 
Now, the twins $3$ and $4$ respectively go once around the hole and  
around the handle of the torus. From a topological point of view,  
their paths can be characterized by a so--called winding index.  In a  
cylinder, the winding index is just an integer which counts the number  
of times a loop rolls around the surface.  In the case of a 2--torus,  
the winding index is a couple $(m,n)$ of integers where $m$ and $n$  
respectively count the numbers of times the loop goes around the hole  
and the handle.  In our example, twins $1$ and $2$ have the same  
winding index (0,0), whereas twins $3$ and $4$ have winding indexes  
respectively equal to $(1,0)$ and $(0,1)$.  The winding index is a  
topological invariant for each traveller: neither change of  
coordinates or reference frame (which ought to be continuous) can  
change its value. 
 
To summarize, we have the two situations: 
\begin{enumerate} 
\item Two twins have the same winding index (twins $1$ and $2$ in our  
example), because their loops belong to the same homotopy class.  
Nevertheless only one (twin $1$) can go from the first meeting point to 
the second one without changing inertial frame. The situation is  
not symmetrical about $1$ and $2$ due to local acceleration, and $1$ is older than $2$. 
\item Several twins ($1$, $3$ and $4$ in our example) can go from the 
first meeting point to a second one at a constant speed, but travel along 
paths with different winding indexes. 
Their situations are not symmetrical because their loops belong to different  
homotopy classes: 
$$\gamma_1\not\sim\gamma_3\not\sim\gamma_4.$$ 
Twin $1$ is older than twins $3$ and $4$ because her path has a zero  
winding index. 
\end{enumerate} 
 
In order to solve exhaustively the twin paradox in a multi--connected  
space, one would like not only to compare separately the ages of the  
travellers with the age of the sedentary twin, but also to compare the  
ages of the various travellers when they respectively meet each other.   
It is clear that the only knowledge of the winding index of their  
loops does not allow us, in general, to compare their various proper  
time lapses.  The only exception is that of the cylinder, where a  
larger winding number always corresponds to a shorter proper time  
lapse.  But for a torus of unequal lengths, for instance when the  
diameter of the hole is much larger than the diameter of the handle, a  
traveller may go straight around the handle many times with a winding  
index $(0,n)$, and yet be older than the traveller which goes straight  
only once around the hole with a winding index (1,0).  The situation  
is still more striking with a double torus, indeed a {\it hyperbolic}  
surface instead of an Euclidean one (see e.g.  \cite{luminet95}).  The  
winding indexes become quadruplets of integers and, as for the simple  
torus, they cannot be  
compared to answer the question on the ages of the travellers.  As we  
shall see below, this problem can be solved only by using an  
additional {\it metric} information.

\section{Langevin and Poincar\'e}\label{IV}

We have found a topological invariant attached to each twin's  
worldline which accounts for the asymmetry between their various  
inertial reference frames.  Why is it so?  In special relativity  
theory, two reference frames are equivalent if there is a Lorentz  
transformation from one frame of space-time coordinates to another  
system.  The set of all Lorentz transformations is called the  
Poincar\'e group -- a ten dimensional group which combines  
translations and homogeneous Lorentz transformations called ``boosts". 
 
The non equivalence between the inertial frames is due to the fact  
that space topology breaks {\it globally} the Poincar\'e group.   
Indeed, cutting and pasting to compactify the spatial sections defines  
(i) particular directions, so that space, even if locally isotropic,  
is no more invariant under rotations, and (ii) a particular time: the  
one measured by an observer whose 4--velocity is perpendicular to  
$\Sigma$ (the twin 1 in our example). Let us call $t$ the proper time  
of such an observer; then the spacetime coordinates $p$ of any point  
$P\in{\cal M}$ can be decomposed in the inertial frame ${\bf K}$  
as $p=(t,{\bf x})$, and the choice of a topology reduces to the choice  
of the identifications 
\begin{equation}\label{identic} 
p=(t,{\bf x})\equiv (t,g{\bf x})=g(p),\quad g\in\Gamma. 
\end{equation} 
In the inertial frame ${\bf K'}$ of the traveller, the coordinates of 
$P$ are given by a Lorentz transformation ${\cal L}(p)=(t',{\bf x}')$ 
with 
\begin{equation}\label{7} 
t'=\gamma\left(t-{\bf v}.{\bf x}\right),\quad 
{\bf x}'=\gamma\left({\bf x}-{\bf v}t\right) 
\end{equation} 
where $\gamma\equiv(1-v^2)^{-1/2}$. It is clear 
from (\ref{identic}) and (\ref{7}) that 
\begin{equation} 
\forall g\not\sim I_d,\, 
\forall p,\quad g\circ{\cal L}(p)\not={\cal L}\circ g(p). 
\end{equation} 
The identification (\ref{identic}) particularises a given foliation  
and spatial sections, leading to the existence of an {\it absolute}  
rest frame (the one of zero homotopy class).  For any other  
observers, these identifications are relations between events at  
different times (and thus in different spatial sections) and not a  
relation between points in a given spatial section.  As pointed in  
\cite{low90}, the observers 3 and 4 will find that their constant time  
hypersurfaces do not match in the universal covering space and that  
there are points on these surfaces of simultaneity which are connected  
by timelike curves.  Moreover, the only holonomy $g$ such that  
$g\circ{\cal L}(p)={\cal L}\circ g(p)$ for all $p$ is the identity  
$g=I_d$, thus the holonomy group reduces to $\Gamma=\lbrace  
I_d\rbrace$ and $\Sigma$ reduces to $X$.  We deduce that the only  
topology compatible with the Poincar\'e group is the trivial topology.   
In other words, the only flat spaceform invariant under the full  
Poincar\'e group is the original simply--connected Minkowski  
spacetime, and any additional discrete identification group is  
incompatible with the Lorentz transformations.  
 
In conclusion, the oldest twin will always be the one of homotopy  
class $\lbrace0\rbrace$, and between two twins of same homotopy class, the oldest  
one will be the one who does not undergo any acceleration.  We can  
rephrase Bondi's formulation of the solution by saying that ``there is  
only one way {\it of a given homotopy class} of getting from the first  
meeting point to the second without acceleration''. 
 
This generalises the previous works 
\cite{brans73,low90,dray90} by adding topological considerations which  
are 
more general and  hold whatever the shape of space is.  As concluded in 
\cite{low90}, ``it is not sufficient that [the] motion [of the twin] is 
symmetrical in terms of acceleration felt and so on; it must also be 
symmetrical in terms of the way that their worldlines are embedded into 
the spacetime''; this latter symmetry is the one we have exhibited and 
which is encoded in the homotopy class.  
 
\begin{figure} 
\centering{  
\epsfig{figure=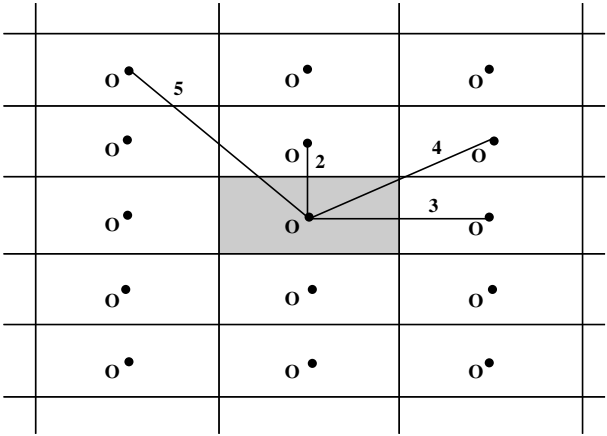, width=8cm}  
\vskip0.25cm  
\caption{Straight paths in the universal covering space of a  
(2+1)--spacetime with flat, torus--like spatial sections.  Paths $2$,  
$3$, $4$, $5$ are straight loops with respective winding indexes  
(0,1), (1,0, (1,1), (1,2), allowing the traveller twins to leave and  
meet again the sedentary twin $O$ without accelerating.  The inertial  
worldlines are clearly not equivalent: the longest  
the spatial length in the universal covering space, the shortest the  
proper time travel in spacetime.} 
\label{fig3}} 
\end{figure}

As mentioned in the previous section, the homotopy classes only tell  
us which twin is aging the fastest: the one who follows a straight  
loop homotopic to $\lbrace0\rbrace$.  It does not provide a classification of the  
ages (i.e.  proper time lengths) along all the straight loops.  To do  
this, some additional information is necessary, such as the various  
identification lengths.  Indeed there exists a simple criterion which  
works in all cases: a shorter spatial length in the universal  
covering space will always correspond to a longest proper time.  To  
fully solve the question, it is therefore sufficient to draw the universal  
covering space as tessellated by the fundamental domains, and to  
measure the lengths of the various straight paths joining the twin  
$1$ position in the fundamental domain to its ghost positions in the  
adjacent domains (see figure~\ref{fig3}).  As usual in topology, all  
reasonings involving  metrical measurements can be solved in the  
simply--connected universal covering space.

In the framework of general relativity, general solutions of  
Einstein's field equations are curved spacetimes admitting no  
particular symmetry.  However, all the exact known solutions admit  
symmetry groups (although less rich than the Poincar\'e group).  For  
instance, the usual ``big bang" cosmological models -- described by  
the Friedmann--Lema\^{\i}tre solutions -- are assumed to be globally  
homogeneous and isotropic.  From a geometrical point of view, this  
means that spacelike slices have constant curvature and that space is  
spherically symmetric about each point.  In the language of group  
theory, the spacetime is invariant under a six-dimensional isometry  
group.  The universal covering spaces of constant curvature are ${\bf  
R}^3$, ${\bf S}^3$ or ${\bf H}^3$ according to the zero, positive or  
negative value of the curvature.  Any identification of points in  
these simply-connected spaces via a holonomy group lowers the  
dimension of their isometry group\footnote{There is one exception:  
the projective space, obtained by identifying the antipodal points  
of ${\bf S}^3$}; it preserves the three--dimensional  
homogeneity group (spacelike slices have still constant curvature),  
but it breaks down globally the isotropy group (at a given point there  
are a discrete set of preferred directions along which the universe  
does not look the same). 
   
Thus in Friedmann--Lema\^{\i}tre universes, (i) the expansion of the  
universe and (ii) the existence of a non--trivial topology for the  
constant time hypersurfaces both break the Poincar\'e invariance and  
single out the same ``privileged" inertial observer who will age more  
quickly than any other twin: the one comoving with the  
cosmic fluid -- although aging more quickly than all her travelling  
sisters may be not a real privilege!

\end{document}